\documentclass{article}
\usepackage{cite}
\usepackage{graphicx}

\begin{document}

\newcommand{\bear}{\begin{eqnarray}}
\newcommand{\eear}{\end{eqnarray}}
\newcommand{\be}{\begin{equation}}
\newcommand{\ee}{\end{equation}}
\newcommand{\beqn}{\begin{eqnarray}}
\newcommand{\eeqn}{\end{eqnarray}}
\newcommand{\beqnn}{\begin{eqnarray*}}
\newcommand{\eeqnn}{\end{eqnarray*}}

\def\vep{\varepsilon}
\def\vf{\varphi}
\def\al{\alpha}

\begin{center} {\Large \bf
Coherent Phase States in the Coordinate and Wigner Representations}

\end{center}

\begin{center} {\bf
Miguel Citeli de Freitas$^1$
and Viktor V. Dodonov$^{1,2}$
}\end{center}
 
\begin{center}

{\it $^1$
Institute of Physics,,
University of Brasilia, 
70910-900 Brasilia, Federal District, Brazil}

$^2$ {International Center for Physics, University of Brasilia, Brasilia, DF, Brazil }

\end{center}

{Correspondence: vdodonov@unb.br}

\abstract{We study numerically the coordinate wave functions and the Wigner functions of the coherent phase states (CPS),
paying the main attention to their differences from the standard (Klauder--Glauber--Sudarshan) coherent states, especially
in the case of high mean values of the number operator. In this case, the CPS can possess a strong coordinate (or momentum)
squeezing, which is, roughly, twice weaker than for the vacuum squeezed states. 
The Robertson--Schr\"odinger invariant uncertainty product in the CPS logarithmically increases with the mean value of the number operator
(whereas it is constant for the standard coherent states). Some measures of (non)Gaussianity of CPS are considered.
}


\section{Introduction}
\label{sec-intr}

The problem of phase in quantum mechanics is as old as quantum mechanics itself 
\cite{London26,Carr68,Garr70,Loudon73,Paul74,Leblond76,Berg91,Popov91,Vogel91,Noh92,Luks94,Lynch95,Dubin95,Tanas96,Royer96,BuHil96,%
Klim97,Pegg97,Perina98,Luis00,Voron02,Kastrup03,Arag04,Varro15}.
One of its main ingredients is the {\em exponential phase operator\/}
\be
E_- = (\hat{n} +1)^{-1/2} \hat{a} , \qquad
\hat{n} = \hat{a}^{\dagger}\hat{a}, \qquad [\hat{a},\hat{a}^{\dagger}] =1.
\label{E}
\ee
It can be found already in the classical paper by Dirac \cite{Dirac27} (although in a slightly incorrect form, 
corrected by London \cite{London27}),
but the systematic study waited until the paper by Susskind and Glogower \cite{Suss64}
(see the history in \cite{Nieto93}).
Various applications and generalizations of operator (\ref{E}) were studied, e.g., in papers 
\cite{Lerner68,Eswaran70,Dod93,Ban95,Oliv03,DoMiDo07,Kozlov17}.

Many researchers studied the properties of eigenstates of the exponential phase operators:
\be
|\vep\rangle = \sqrt{1-|\vep|^2} \sum_{n=0}^{\infty} \vep^n |n\rangle, \qquad \vep = |\vep|e^{i\vf}, \quad |\vep| < 1, \quad
\hat{a}^{\dagger}\hat{a} |n\rangle = n |n\rangle.
\label{defCPS}
\ee
These states were introduced in papers \cite{Lerner70,Ifantis72,Ahar73} without any special name. 
The name ``coherent phase states'' was coined by Shapiro and Shepard \cite{ShapShep91}.
The name ``harmonious state'' was used by Sudarshan \cite{Sud93}. The name ``pseudothermal states'' was suggested in Ref. \cite{DoMi95},
because the state (\ref{defCPS}) can be considered as a pure-state analog of the mixed equilibrium state of the harmonic oscillator,
described by means of the statistical operator
\be
\hat\rho_{th} = \left(1-|\vep|^2\right) \sum_{n=0}^{\infty} |\vep|^{2n} |n\rangle \langle n|,
\label{rhoth}
\ee
with the mean number of quanta
\be
\langle \hat{n}\rangle_{\vep} = \langle \hat{n}\rangle_{\rho} = 
\left(1-|\vep|^2\right)|\vep|^{2} \sum_{n=0}^{\infty} |\vep|^{2n}(n+1) =
\frac{|\vep|^2}{1- |\vep|^2}, \qquad
|\vep|^2 = \frac{\langle \hat{n}\rangle}{1 + \langle \hat{n}\rangle}.
\label{meann}
\ee
The states (\ref{defCPS}) [named sometimes as ``phase coherent states''] 
were also considered from different points of view in Refs. 
\cite{Vourd90,Agarw91,Vourd92,Hall93,Vourd96,Obada97,Paris99,DoMi02,Gerry09,Luis10,Luis11a,Luis11,Dod11,Wun15}.
In particular, the physical meaning of these states and their distinguished property was clarified in paper \cite{DAriano96}:
``phase-coherent states preserve their coherence under amplification and achieve the best amplifier performance''.
More recently, it was shown in paper \cite{Becir12} that ``the use of phase coherent states gives better performance in terms of noise tolerance, key rate, and achievable distance than that of common linear coherent states'' 
for continuous variable quantum key distribution protocols.
Probably, such states can be used to understand the quantum phase transition in the squeezed Jaynes--Cummings model \cite{Shen22}.
Methods of generation of the states (\ref{defCPS}) were proposed in \cite{DAriano98,Baseia98}.
Further generalizations were studied, e.g., in Refs. \cite{Brif95,Chizhov98,Wunsche01,Wun03,Mouayn12,Jagan20}.
In particular, it appears \cite{Jagan20,Sivak00,Obada04} that the coherent phase states are the special case of a large family of
{\em nonlinear coherent states\/} introduced in papers \cite{Vogel96,Man-nonlin97}.

In the papers cited above, the authors studied the properties of state (\ref{defCPS}) from the point of view of
relations between the {\em phase and number variables}. Our goal is to study the properties of state (\ref{defCPS})
with respect to the {\em canonical quadrature operators\/} 
\[
 \hat{x} = \left( \hat{a} +  \hat{a}^{\dagger}\right)/\sqrt{2}, \quad
  \hat{p} = \left( \hat{a} -  \hat{a}^{\dagger}\right)/(i\sqrt{2}),
  \]  
comparing these properties with that of the standard (Klauder--Glauber--Sudarshan)  \cite{Klaud60,Glaub63,Sudar63}
coherent states
\be
|\alpha\rangle = \exp\left(-|\alpha|^2/2\right)\sum_{n=0}^{\infty} \frac{\alpha^n}{\sqrt{n!}}|n\rangle.
\label{coh}
\ee
It is well known that the wave function of state (\ref{coh}) in the coordinate representation has the Gaussian form 
\cite{Schrod26} with equal constant variances 
$\sigma_x^{\alpha} = \sigma_p^{\alpha} = 1/2$:
\be
\langle x|\alpha\rangle = \pi^{-1/4} \exp\left(-\frac12 x^2 +\sqrt{2}\,x\alpha -\frac12 \alpha^2 -\frac12 |\alpha|^2 \right).
\label{psicoh}
\ee
We wish to know, how the absence of the denominator $\sqrt{n!}$ in the  expansion (\ref{defCPS})
influences the mean values and variances of the coordinate and momentum operators, as well as the form of the wave function.
In view of the equidistant spectrum of the harmonic oscillator, the time evolution of state (\ref{defCPS}) is equivalent to
the linear increase of the phase $\vf(t) = \vf(0) + t$ (formally, we consider the situation when $\omega=m=\hbar =1$).

\section{Quadrature mean values, variances and the Robertson -- Schr\"odinger uncertainty product }
\label{sec-var}

Mean values of the quadrature operators in the state (\ref{defCPS}) have the form
\be
\langle \hat{x}\rangle = \sqrt{2}\,\cos(\vf) S_1, \qquad \langle \hat{p}\rangle = \sqrt{2}\,\sin(\vf) S_1,
\ee
\be
S_1 = \left(1-|\vep|^2\right) \sum_{n=0}^{\infty} |\vep|^{2n+1}\sqrt{n+1} 
=(1 + \langle \hat{n}\rangle)^{-1} \sum_{n=0}^{\infty} 
\left(\frac{\langle \hat{n}\rangle}{1 + \langle \hat{n}\rangle}\right)^{n+1/2}\sqrt{n+1}.
\label{S1}
\ee
The quadrature (co)variances can be written as
\be
\left.
\begin{array}{c}
\sigma_x \\
\sigma_p
\end{array} \right\}
= N - S_1^2 \pm \left(S_2 -S_1^2\right) \cos(2\vf) , \qquad 
 \sigma_{xp} = \sin(2\vf)\left(S_2 - S_1^2\right),
\ee
\be
S_2 = \left(1-|\vep|^2\right) \sum_{n=0}^{\infty} |\vep|^{2n+2}\sqrt{(n+1)(n+2)}
=(1 + \langle \hat{n}\rangle)^{-1} \sum_{n=0}^{\infty} 
\left(\frac{\langle \hat{n}\rangle}{1 + \langle \hat{n}\rangle}\right)^{n+1}\sqrt{(n+1)(n+2)}.
\ee
Note that for the (mixed) thermal equilibrium states one has
\be
\sigma_x^{therm} = \sigma_p^{therm} = \langle \hat{n}\rangle +1/2 \equiv N = \frac{1+|\vep|^2}{2\left(1-|\vep|^2\right)}, \quad
\sigma_{xp}^{therm} = 0.
\ee

If $\vf = \pi/2$, then $\sigma_x^{\vf=\pi/2} = N - S_2$. Noticing that
\[
\langle \hat{n}\rangle = \left(1-|\vep|^2\right) \sum_{n=1}^{\infty} |\vep|^{2n} n =
\left(1-|\vep|^2\right) \sum_{m=0}^{\infty} |\vep|^{2m+2} (m+1),
\]
we arrive at the formula
\be
\sigma_x^{\vf=\pi/2} =  \frac12 - \left(1-|\vep|^2\right)|\vep|^{2} 
\sum_{n=0}^{\infty} \frac{|\vep|^{2n}\sqrt{n+1}}{\sqrt{n+1} + \sqrt{n+2}} \, .
\label{sigmin}
\ee
Hence, there is squeezing of the $x$-quadrature at $\vf =\pi/2$ 
(and the same degree of squeezing of the $p$-quadrature at $\vf =0$)
for any value of $|\vep|$. 
The existence of squeezing in the coherent phase states was discovered numerically (presented in the form of tables
for $|\vep|^2$ from $0.1$ to $0.9999$ or $\langle \hat{n}\rangle$ from $0.11$ to $9999$) in Ref. \cite{Marhic90}.
For $|\vep| \ll 1$ we have 
$\sigma_x^{\vf=\pi/2} \approx 1/2 - |\vep|^2(\sqrt{2} -1)$.
An ideal squeezing ($\sigma_x^{\vf=\pi/2} \to 0$) can be obtained for $|\vep| \to 1$.

Numerical calculations show that the variance $\sigma_x^{\vf=\pi/2}$ as function of $|\vep|^2$ is close to the straight line
 $\frac12\left(1-|\vep|^2\right)$.
However, the exact function is always above this straight line. 
A good analytic approximation can be found in the asymptotic region $|\vep|^2 \to 1$, if one takes into account that the main contribution to 
the series in Equation (\ref{sigmin}) is made by terms with big values of the summation index $n$. Therefore, we make the approximation
\[
\frac{\sqrt{n+1}}{\sqrt{n+1} + \sqrt{n+2}} = \left(1 +\sqrt{\frac{n+2}{n+1}}\right)^{-1} \approx
\left(2 + \frac1{2(n+1)}\right)^{-1} \approx \frac12\left(1 -\frac1{4(n+1)}\right).
\]
Then, using two exact series,
\[
\sum_{n=0}^{\infty} x^n = (1-x)^{-1}, \qquad \sum_{n=0}^{\infty} \frac{x^{n+1}}{n+1} = \int_0^x dy\sum_{n=0}^{\infty} y^n  = 
-\ln(1-x),
\]
we arrive at the approximate formulas
\be
\sigma_x^{\vf=\pi/2} \approx \frac12\left(1-|\vep|^2\right) \left[ 1 - \frac14 \ln\left(1-|\vep|^2\right)\right]
= \frac1{2\left(1 + \langle \hat{n}\rangle\right)}\left[ 1 + \frac14 \ln\left(1 + \langle \hat{n}\rangle\right)\right].
\label{sigminapr}
\ee
\begin{figure}[hbt]
\centering
\includegraphics[scale=0.11]{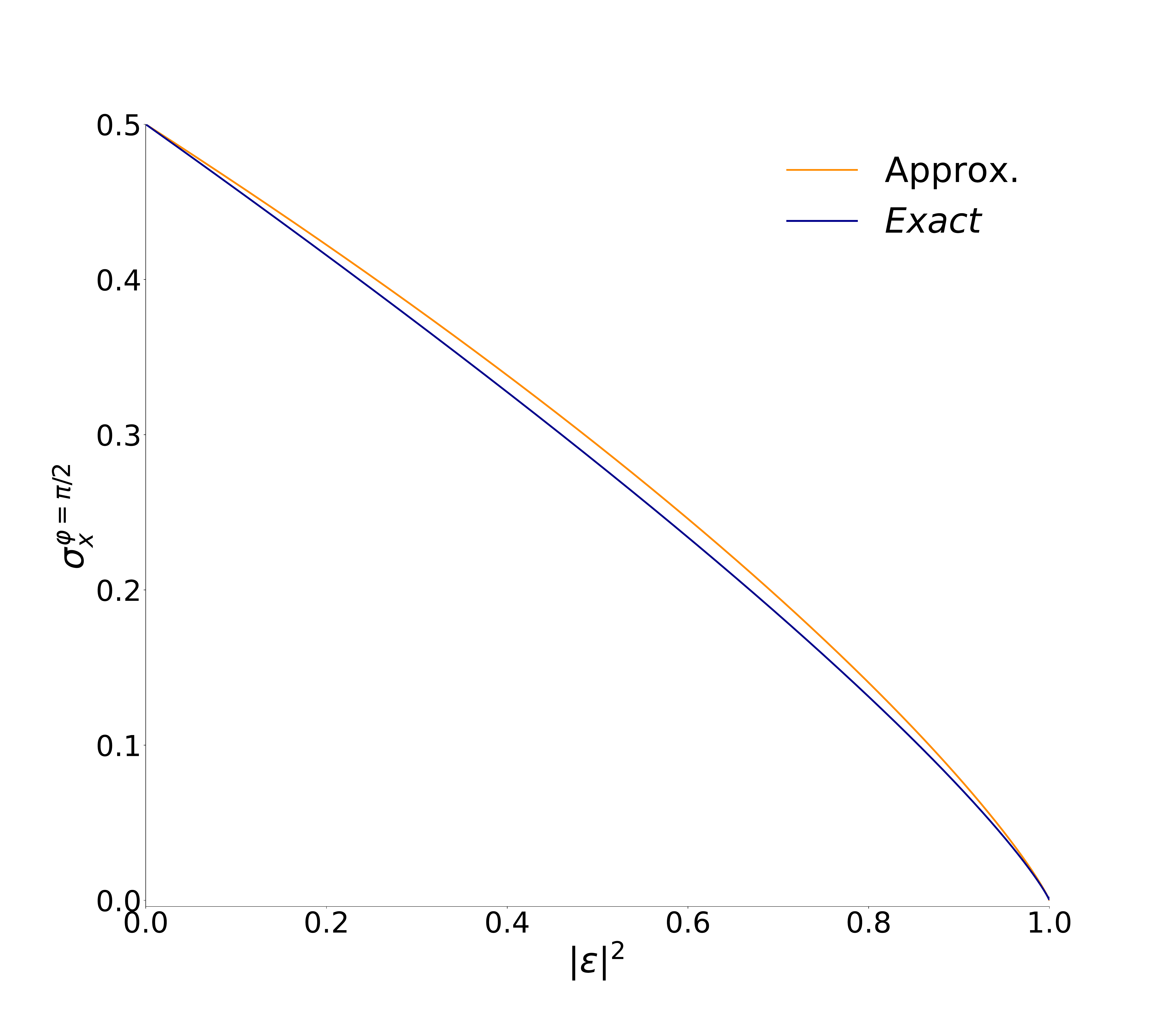}
\hfill
\includegraphics[scale=0.11]{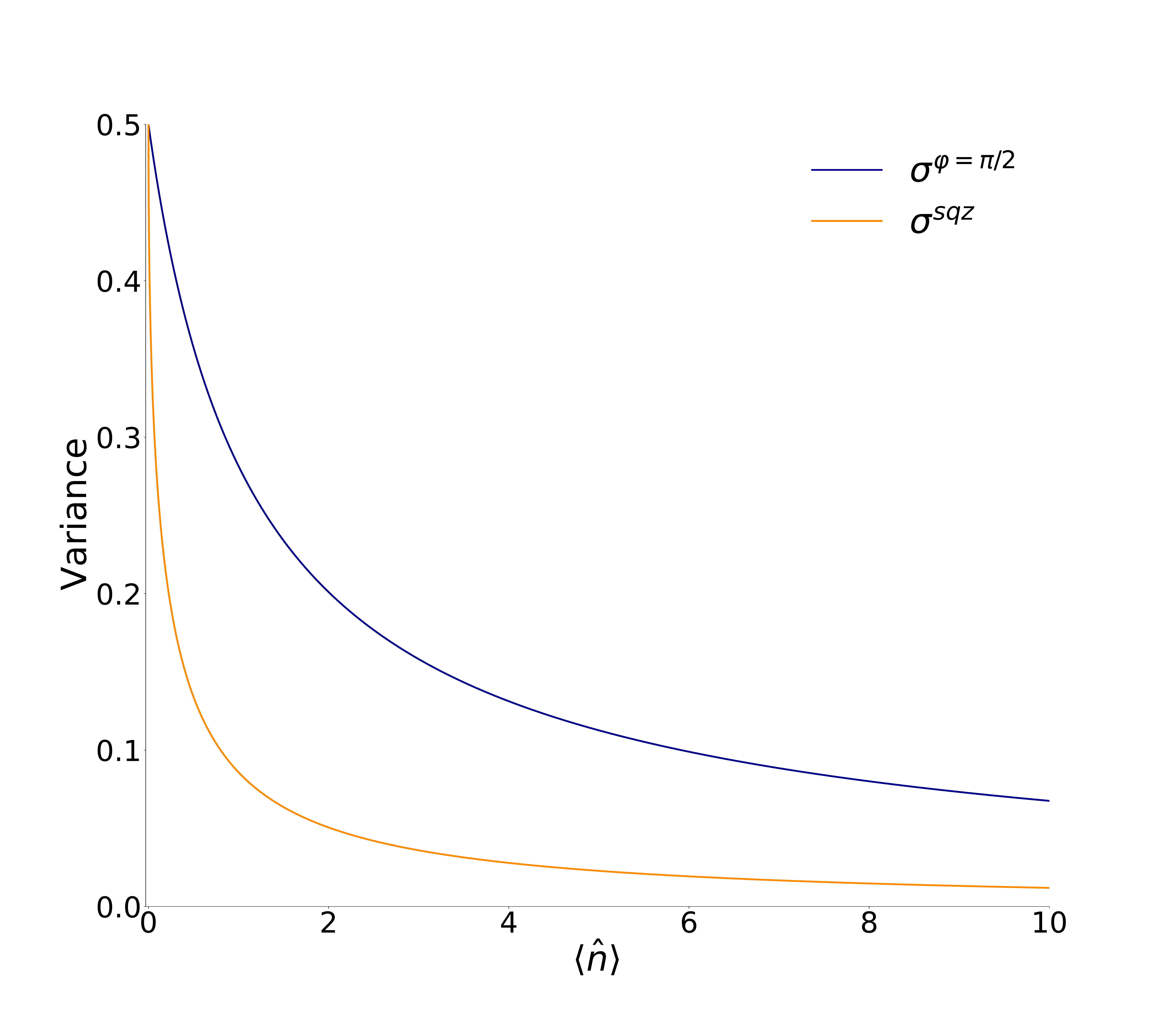}
\caption{\small{Left: The variance $\sigma_x^{\vf=\pi/2}$ versus $|\vep|^2$. 
Exact numeric calculations were made taking into account $1000$ terms in formula (\ref{sigmin}). 
Right: The exact variance $\sigma_x^{\vf=\pi/2}$ versus the mean number of quanta $\langle \hat{n}\rangle$, compared with the
variance (\ref{sqzvac}) in the squeezed vacuum state.
}}
\label{fig-sigmin}
\end{figure} 

The left-hand part of Figure \ref{fig-sigmin} shows $\sigma_x^{\vf=\pi/2}$ as function of  $|\vep|^2$, comparing the exact (numeric)
values with the approximate formula (\ref{sigminapr}). One can see that approximate formula works rather satisfactory for all values 
of $|\vep|^2$.
The right-hand part of Figure \ref{fig-sigmin} shows 
the exact  values of $\sigma_x^{\vf=\pi/2}$
as function of the mean number of quanta $\langle \hat{n}\rangle$.
This dependence is compared 
 with the known formula for the pure squeezed vacuum (Gaussian) state 
with the variance $\sigma_x = (1/2) e^{-2r}$ and $\langle \hat{n}\rangle = \sinh^2(r)$:
\be
\sigma_x^{sqz} = \left\{2\left[2\langle \hat{n}\rangle +1 
+ 2\sqrt{\langle \hat{n}\rangle\left(\langle \hat{n}\rangle+1\right)}\right]\right\}^{-1}.
\label{sqzvac}
\ee
This quantity behaves as $\left[4\left(1 + 2\langle \hat{n}\rangle\right)\right]^{-1}$ for $\langle \hat{n}\rangle \gg 1$.
Consequently, roughly speaking, the squeezing effect in the squeezed vacuum states is twice more strong than in the coherent
phase states.

\begin{figure}[hbt]
\centering
\includegraphics[scale=0.11]{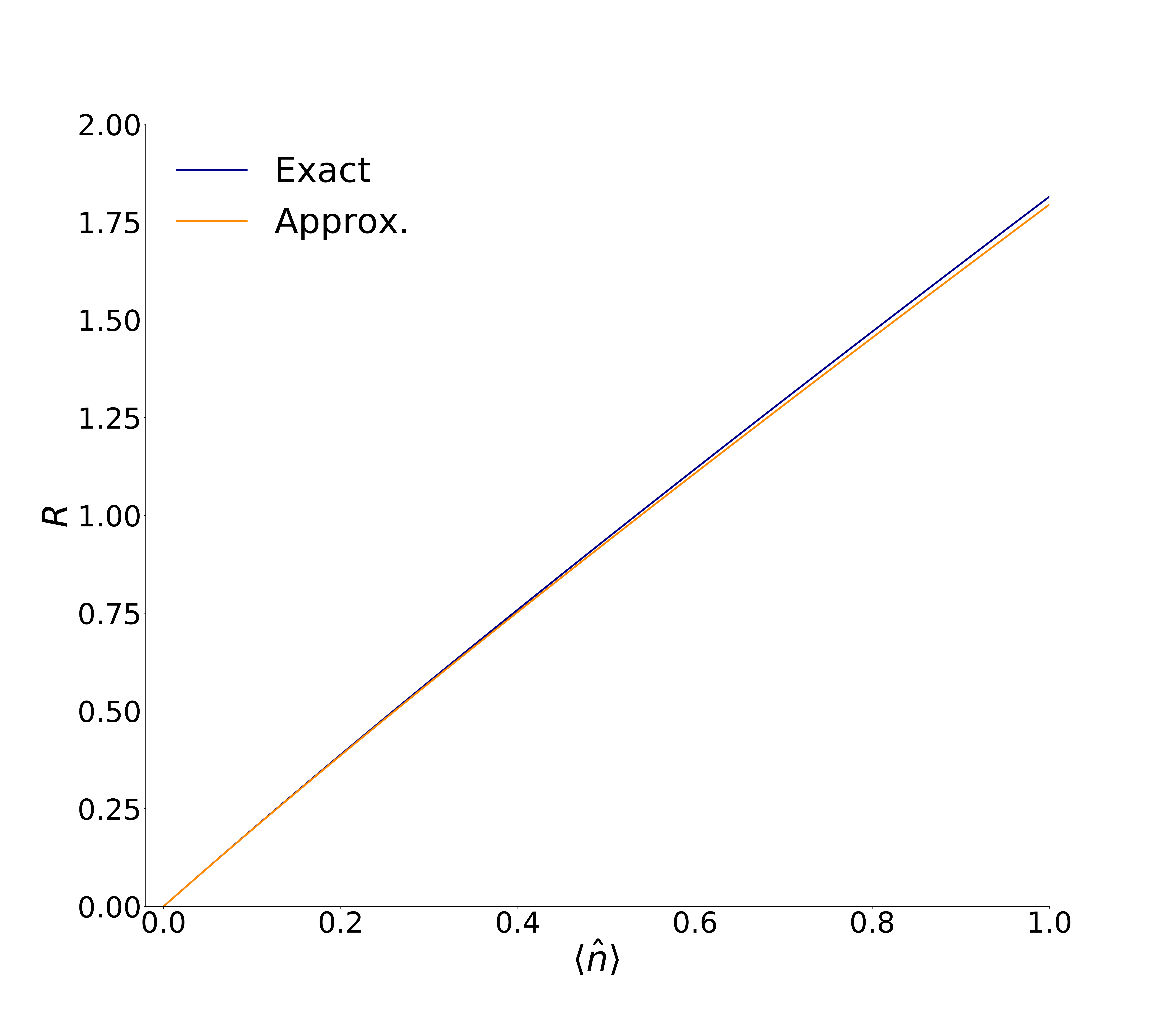}
\hfill
\includegraphics[scale=0.11]{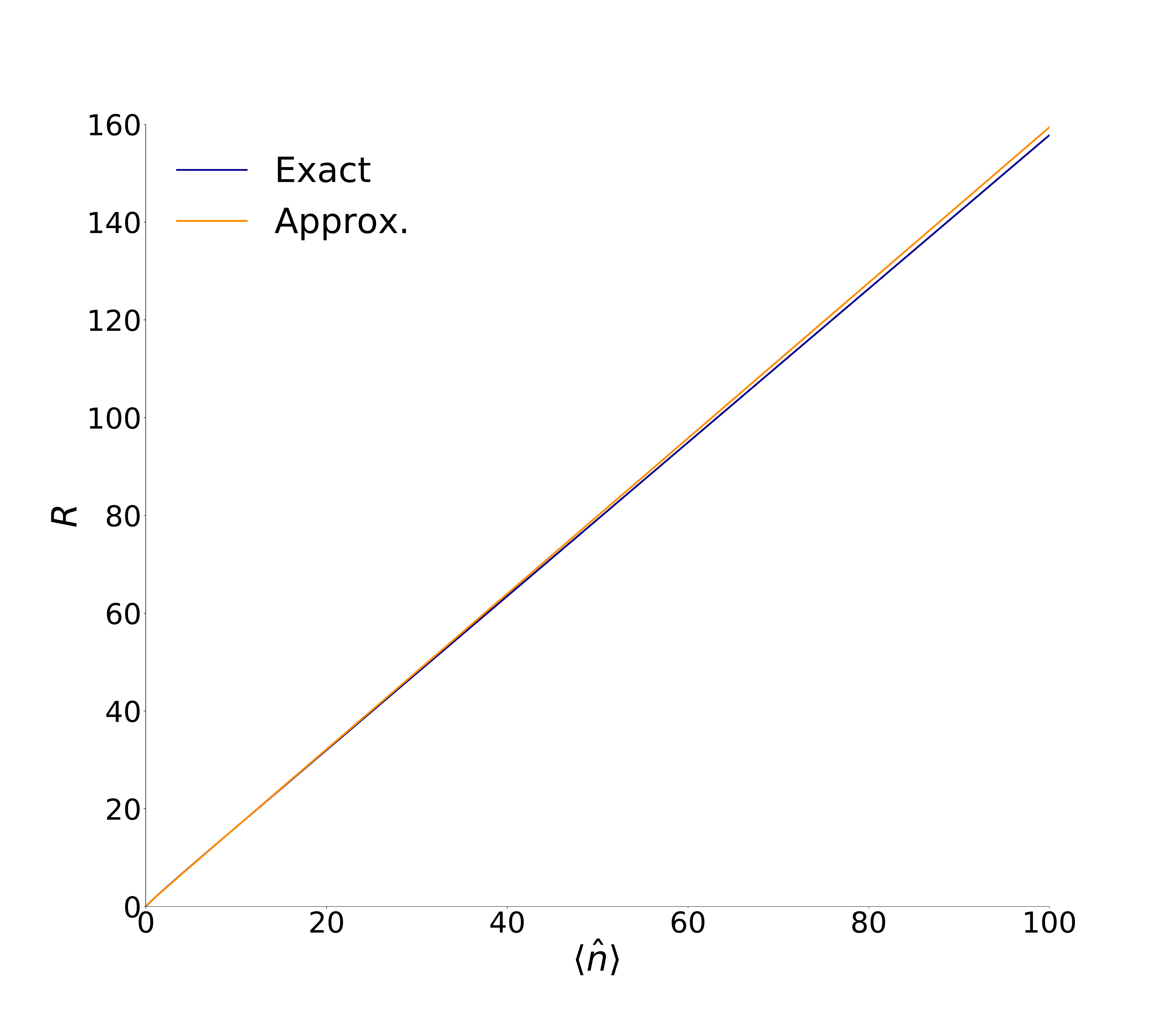}
\caption{\small{Parameter $R$ versus the mean number of quanta $\langle \hat{n}\rangle$.
Left: for $\langle \hat{n}\rangle <1$. 
Right: for $\langle \hat{n}\rangle <100$.
Exact numeric calculations were made taking into account $10000$ terms in formula (\ref{S1}). 
}}
\label{fig-R}
\end{figure}

The mean values of the coordinate and momentum in the coherent state (\ref{coh}) satisfy the relation 
$ R \equiv \langle \hat{x} \rangle^2 + \langle \hat{p} \rangle^2 = 2|\alpha|^2 = 2\langle \hat{n}\rangle$.
For the coherent phase state (\ref{defCPS}) we have $R= 2 S_1^2$, where $S_1$ is given by 
Equation (\ref{S1}).
According to this equation, $R \approx 2\langle \hat{n}\rangle$ for $\langle \hat{n}\rangle \ll 1$. 
However, numeric calculations show that  the plot of function $R(\langle \hat{n}\rangle)$ 
for $\langle \hat{n}\rangle \gg 1$
turns out very close to the straight line with a smaller inclination coefficient:
$R \approx \eta \langle \hat{n}\rangle$  with $\eta \approx 1.59$. 
A reasonable interpolation formula between two regimes can be written as
\be
R \approx \frac{\langle \hat{n}\rangle(2 + \eta \langle \hat{n}\rangle)}{1+\langle \hat{n}\rangle}.
\label{Rappr}
\ee
A comparison of this formula with exact numerical values is shown in Figure \ref{fig-R}

An interesting quantity is the {\em Robertson--Schr\"odinger uncertainty product} \cite{Schrod30,Robertson30}
\be
D \equiv \sigma_x \sigma_p - \sigma_{xp}^2.
\label{RSD}
\ee 
It cannot be smaller than $\hbar^2/4$ (or $1/4$ in the dimensionless units used in this paper).
The equality $D=1/4$ holds for all Gaussian {\em pure\/} states, including the standard coherent states. On the other hand,
$D= (\langle \hat{n}\rangle +1/2)^2$ for the (mixed Guassian) thermal states.
For the coherent phase state (\ref{defCPS}) we have
\be
D = \left(N - S_1^2\right)^2 - \left(S_2 - S_1^2\right)^2
=
\left(N -S_2\right)\left(N +S_2 - 2 S_1^2 \right).
\label{D}
\ee
The phase independence of the right-hand side of Equation (\ref{D}) becomes obvious, 
if one takes into account the equivalence 
between the phase change and time evolution. The quantity $D$ is the {\em quantum universal invariant}, which preserves its value
during the time evolution governed by {\em any\/} quadratic Hamiltonian (in one dimension) \cite{Dod-inv,DOVM-inv}.

\begin{figure}[hbt]
\centering
\includegraphics[scale=0.1]{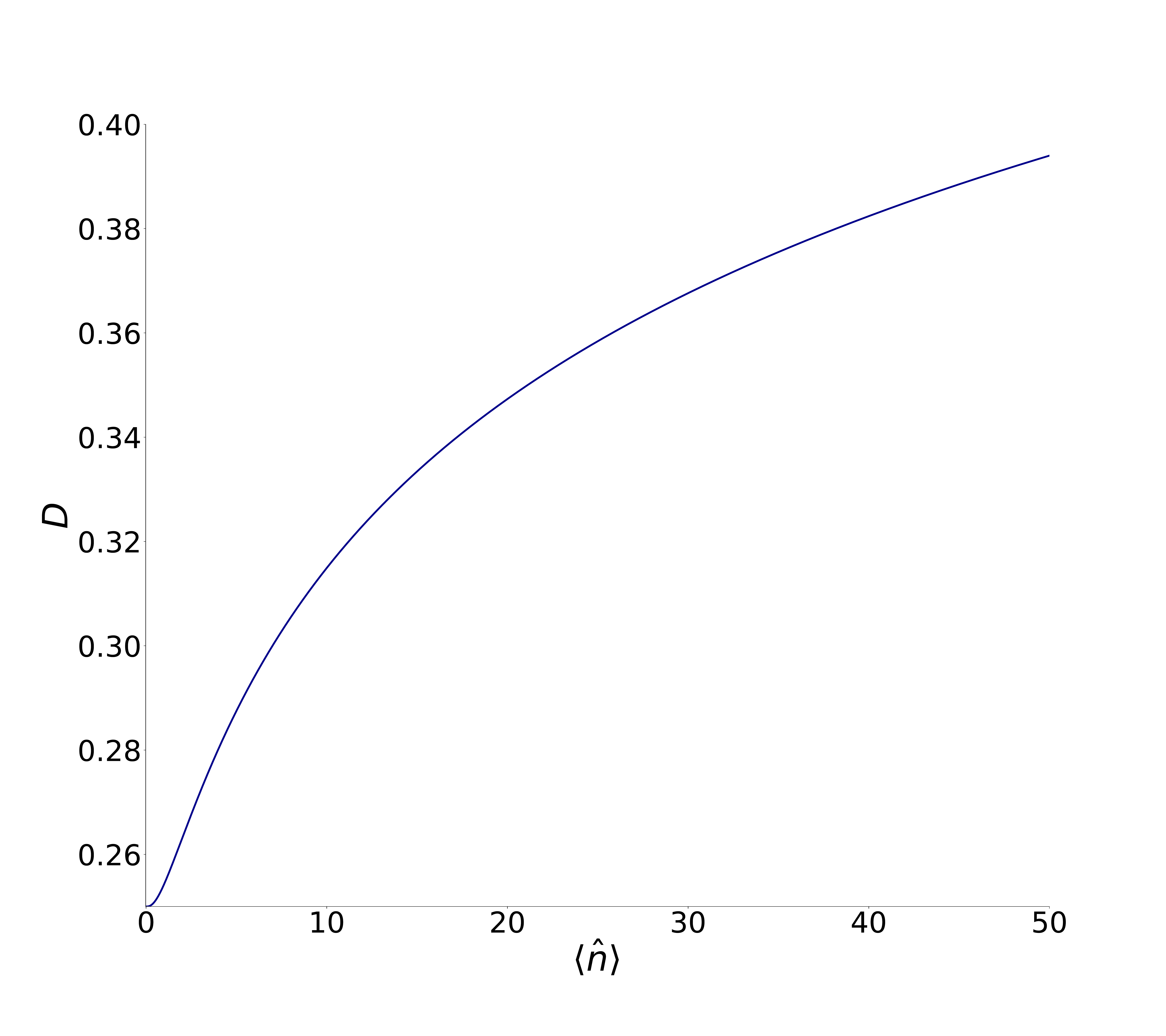}
\hfill
\includegraphics[scale=0.1]{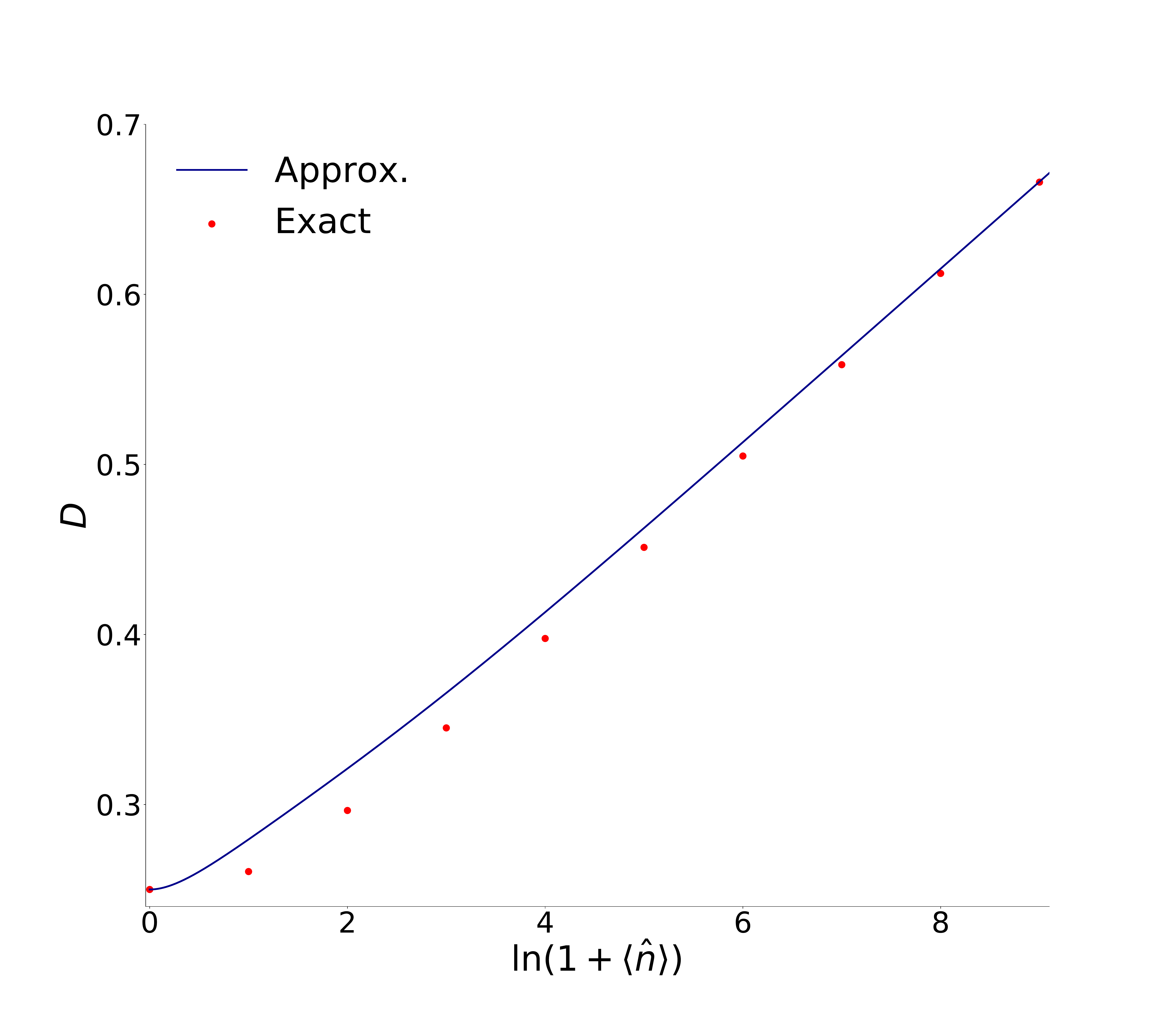}
\caption{\small{The Robertson--Schr\"odinger uncertainty product $D$ versus the mean number of quanta $\langle \hat{n}\rangle$
in the coherent phase state (\ref{defCPS}). 
Left: $D$ as function of  $\langle \hat{n}\rangle$ for relatively small values of $\langle \hat{n}\rangle$ (numeric results).
Right: the comparison of the approximate analytical formula (\ref{Das-1}) with exact numerical values.
The best coincidence happens for $\eta= 1.59$. 
The series $S_1$ and $S_2$ were calculated numerically, taking into account up to $640000$ terms for high values of 
$\langle \hat{n}\rangle$, in order to obtain a good precision.
}}
\label{fig-D}
\end{figure}

The right-hand side of Equation (\ref{D}) can be also written as
\be
D= \sigma_x^{\vf=\pi/2}\left( 2N - \sigma_x^{\vf=\pi/2} - R \right).
\ee
Then, using approximate formulas (\ref{sigminapr}) and (\ref{Rappr}), we arrive at the approximate expression
\be
D  \approx
 \frac{1 +(1/4)\ln\left(1 + \langle \hat{n}\rangle\right)}{4\left(1 + \langle \hat{n}\rangle\right)^2}
\left[ 2(2-\eta)\langle \hat{n}\rangle^2 + 1 + 2\langle \hat{n}\rangle -(1/4)\ln\left(1 + \langle \hat{n}\rangle\right)
\right].
\label{Das-1}
\ee
This formula yields the correct value $D=1/4$ even for $\langle \hat{n}\rangle=0$.
A simplified version for $\langle \hat{n}\rangle \gg 1$,
\be
D \approx \frac{2-\eta}{2}\left[1 + \frac14\ln\left(1 + \langle \hat{n}\rangle\right)\right]
= \frac{2-\eta}{2}\left[1 - \frac14\ln\left(1 -|\vep|^2\right)\right],
\label{Das}
\ee
 yields $D \approx 0.677$ for $1 +\langle \hat{n}\rangle = 10^4$ and $\eta = 1.59$.
The result of Ref. \cite{Marhic90} for $\langle \hat{n}\rangle = 9999$ can be transformed to the form
$D = 0.67$. The dependence of $D$ on $\langle \hat{n}\rangle$ is illustrated in Figure \ref{fig-D}.
\begin{figure}[hbt]
\centering
\includegraphics[scale=0.1]{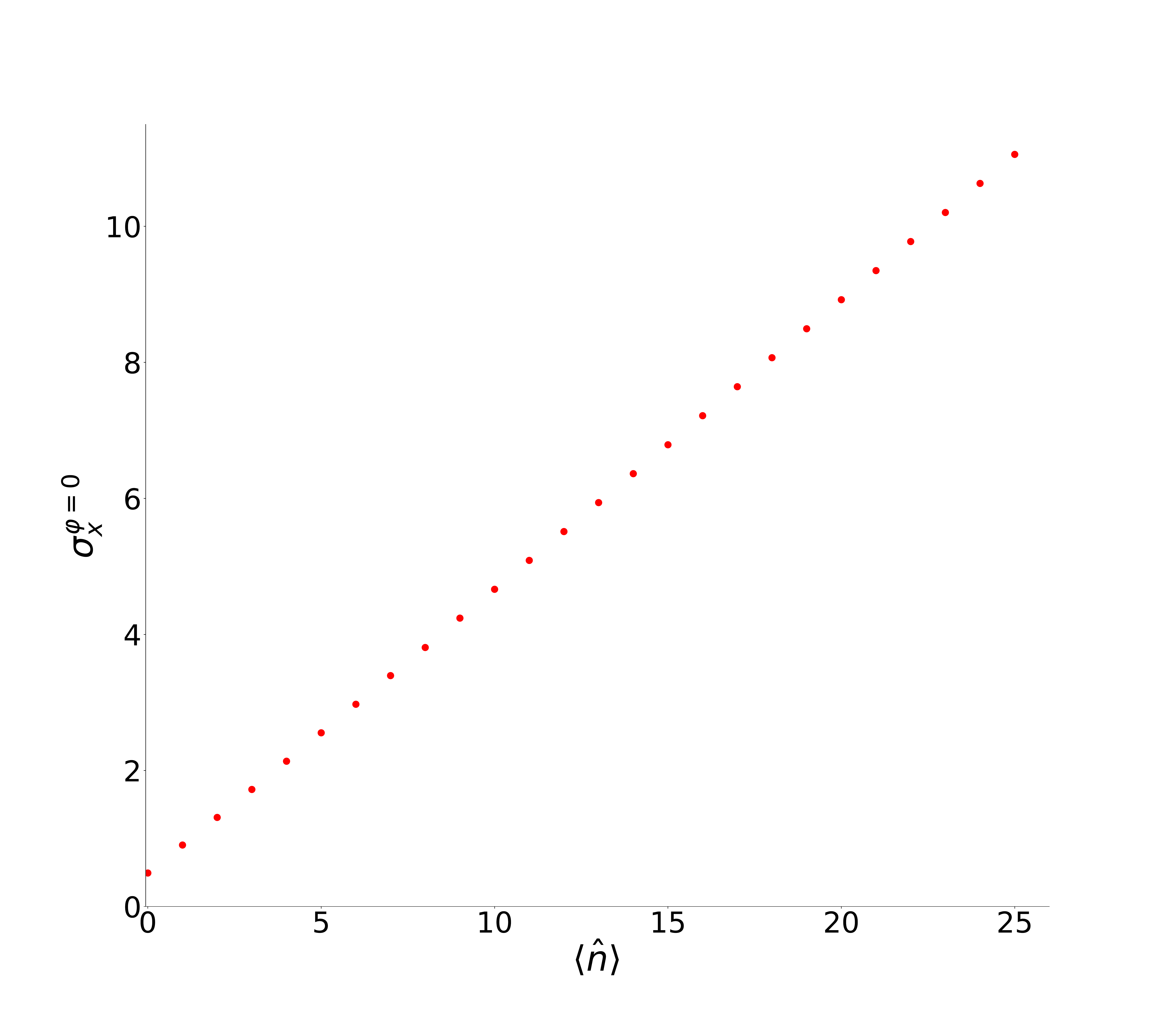}
\hfill
\includegraphics[scale=0.1]{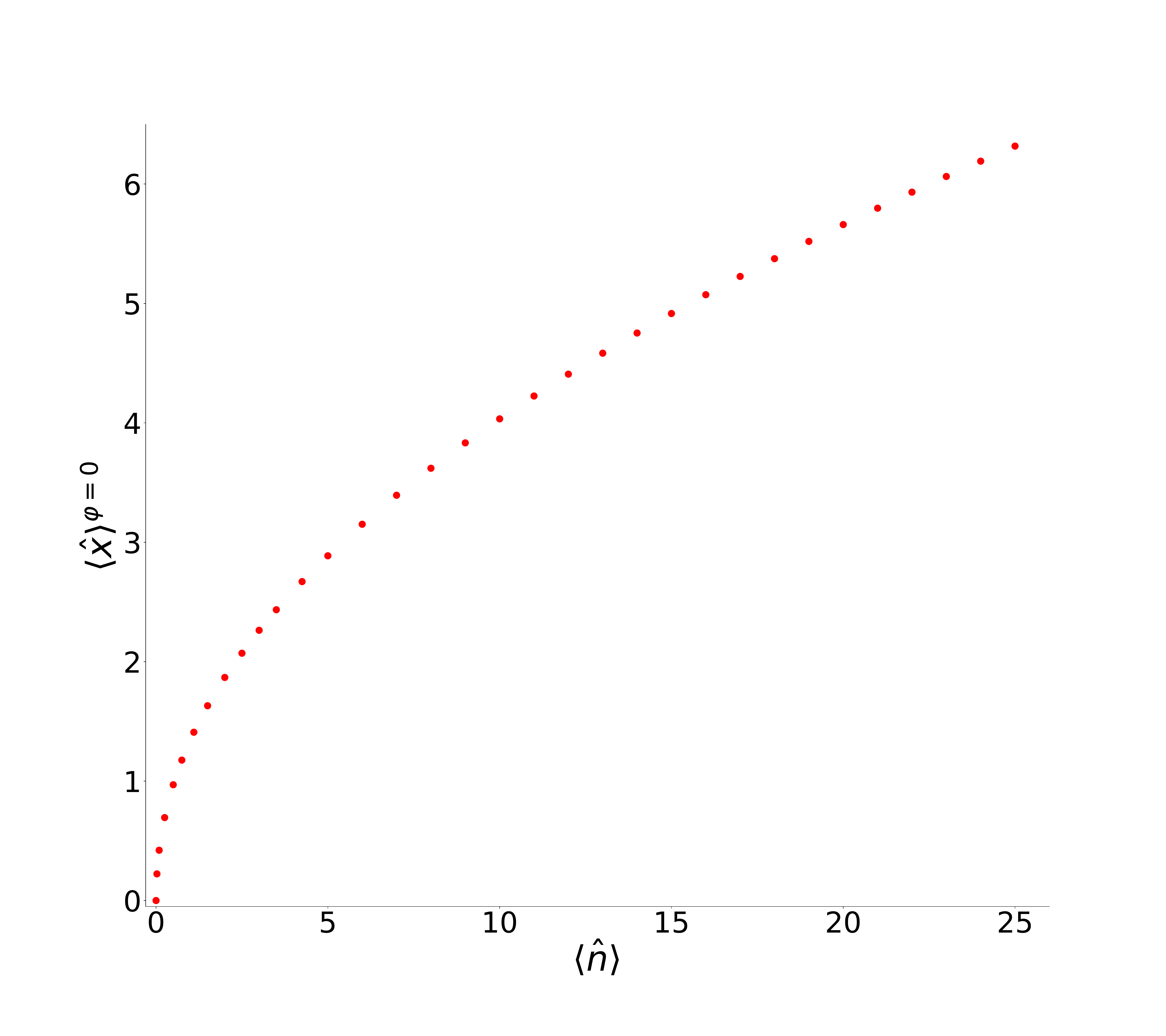}
\caption{\small{Left: The coordinate variance $\sigma_x^{\vf=0}$ versus the mean number of quanta $\langle \hat{n}\rangle$ 
in the coherent phase state (\ref{defCPS}) (numeric results).
Right: The mean value $\langle x\rangle^{\vf=0}$ in the coherent phase state (\ref{defCPS}) (numeric results).
All numeric results were obtained taking into account $10000$ terms in series $S_1$ and $S_2$.
}}
\label{fig-sqz-mean-phi0}
\end{figure} 
Using Eqs. (\ref{sigminapr}), (\ref{D}) and (\ref{Das-1}), and remembering that $\sigma_p^{\vf=0} = \sigma_x^{\vf=\pi/2}$,
 we obtain an approximate formula for the coordinate variance for $\vf=0$:
\be
\sigma_x^{\vf=0} \approx \frac
{2(2-\eta)\langle \hat{n}\rangle^2 + 1 + 2\langle \hat{n}\rangle -(1/4)\ln\left(1 + \langle \hat{n}\rangle\right)}
{2(1 + \langle \hat{n}\rangle)} .
\label{sigx0}
\ee
If $\langle \hat{n}\rangle \gg 1$, $\sigma_x^{\vf=0} \approx (2-\eta) \langle \hat{n}\rangle \approx 0.4 \langle \hat{n}\rangle$.
Exact numeric values of $\sigma_x^{\vf=0}$ and $\langle x\rangle^{\vf=0}$ are plotted in Fig.~\ref{fig-sqz-mean-phi0}.
Remember that in the standard coherent state (\ref{coh}) with a real parameter $\alpha$ we have 
$\langle x\rangle = \sqrt{2}\langle \hat{n}\rangle$, while $\sigma_x =1/2 =const$.

\section{Wave function and probability density}

The complex wave function $\psi_{\vep}(x) \equiv \langle x|\vep\rangle$ is the infinite sum of the Hermite polynomials:
\be
\left(
\begin{array}{c}
\mbox{Re}\psi_{\vep}(x)
\\
\mbox{Im}\psi_{\vep}(x)
\end{array}
\right) = \pi^{-1/4} \sqrt{1-|\vep|^2}\exp(-x^2/2) \sum_{n=0}^{\infty} \frac{ |\vep|^n H_n(x)}{\sqrt{2^n n!}} \times
\left(
\begin{array}{c}
\cos(n\vf)
\\
\sin(n\vf)
\end{array}
\right).
\label{psi-gen}
\ee
The most interesting case is $\langle \hat{n}\rangle \gg 1$. Using the results of the preceding section, we can expect that for $\vf=0$
the wave function is real and very wide, with the maximum near the point $x_* \approx \sqrt{\eta \langle \hat{n}\rangle}$
and the width of the order of $\sqrt{\sigma_x} \sim \sqrt{D/\sigma_p^{\vf=0}} \sim \sqrt{\langle \hat{n}\rangle}$
(so that the maximal value of $\psi(x)$ is of the order of $\langle \hat{n}\rangle^{-1/4}$). 
On the contrary, if $\vf=\pi/2$, the probability density $|\psi_{\vep}(x)|^2$ is concentrated nearby the point $x=0$ in the very
narrow region, whose width is of the order of $(2\langle \hat{n}\rangle)^{-1/2}$. Hence, we can expect that the height of the
wave function in this case is of the order of $\langle \hat{n}\rangle^{1/4}$. 

\begin{figure}[hbt]
\centering
\includegraphics[scale=0.13]{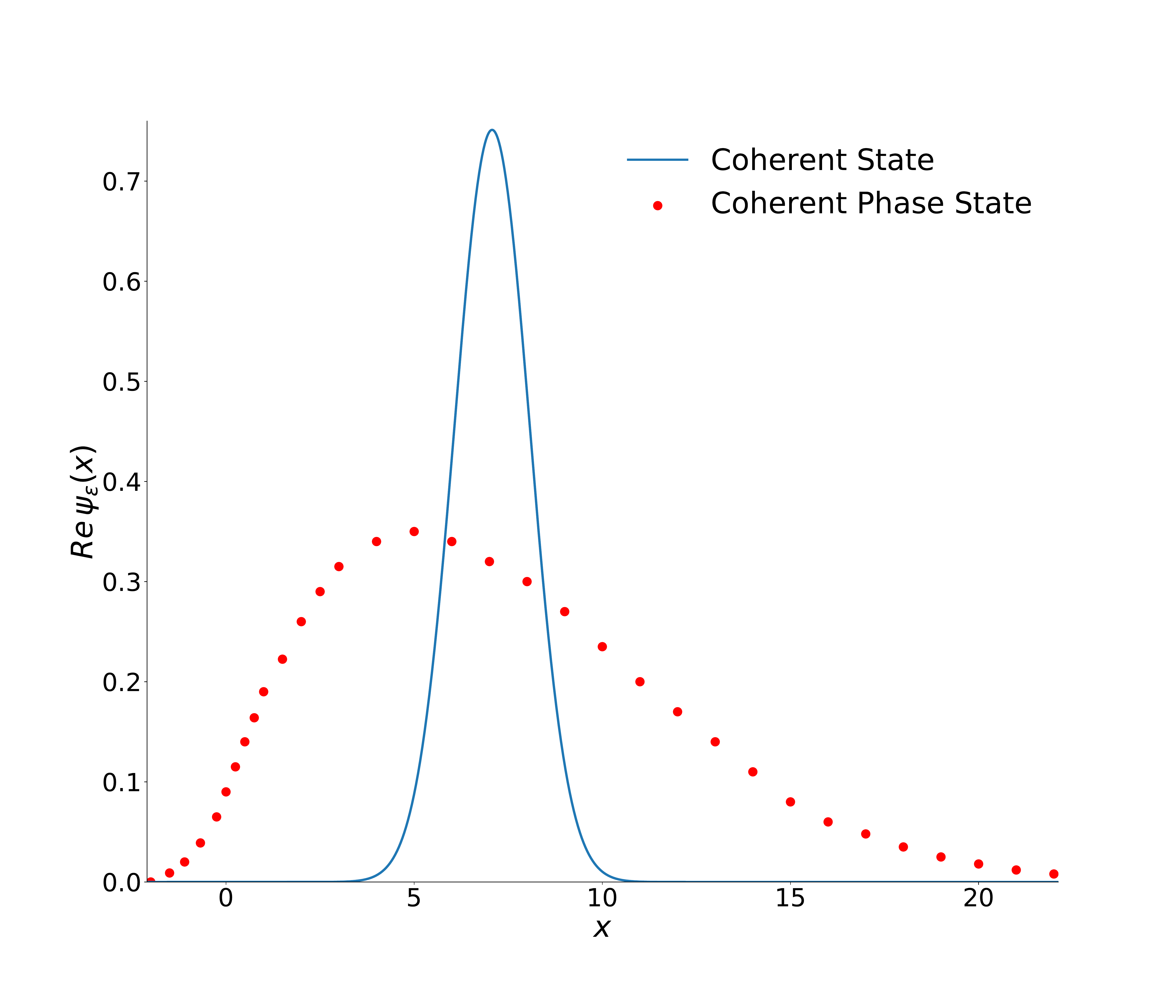}
\hfill
\includegraphics[scale=0.10]{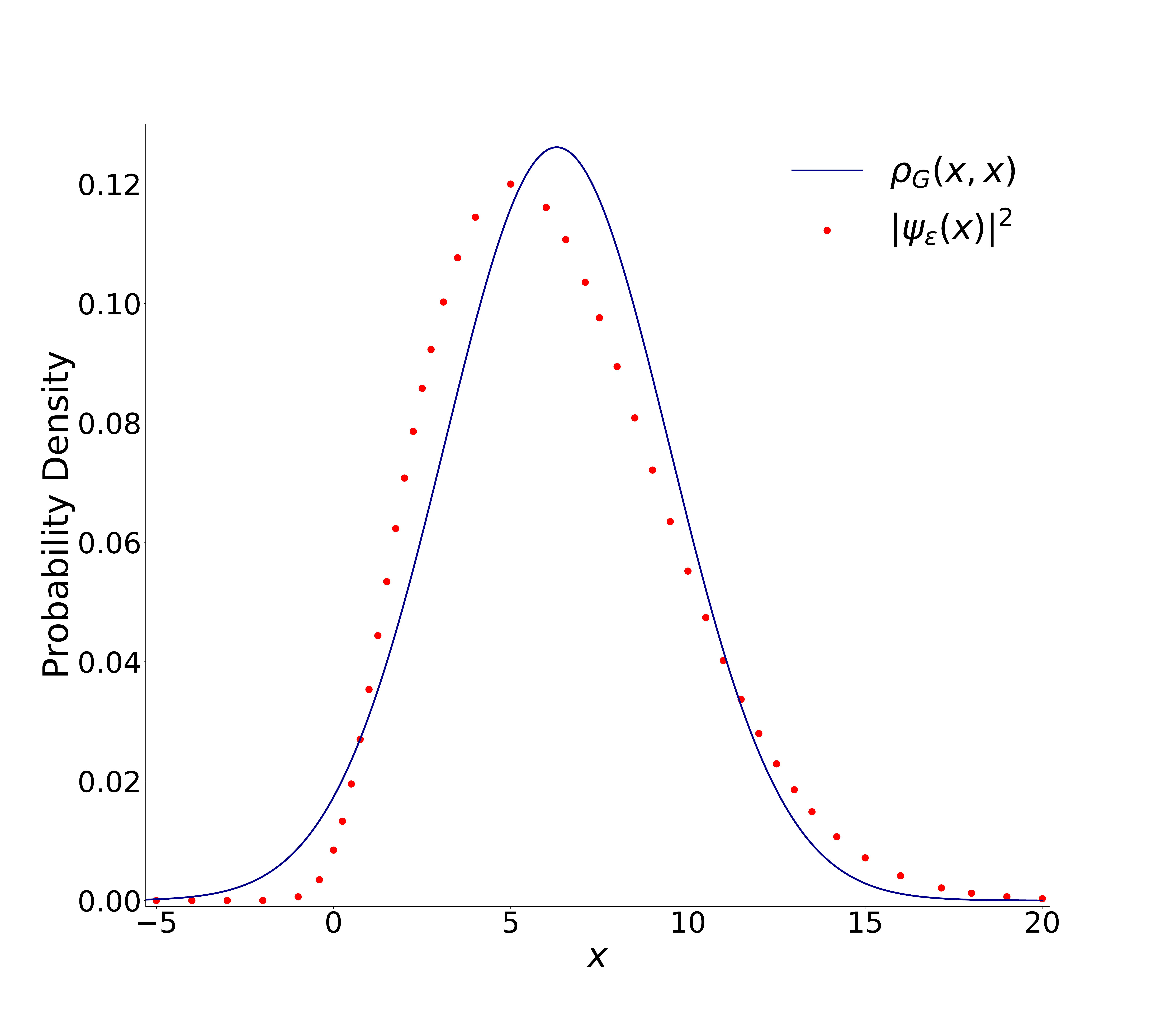}
\caption{\small{Left: The real wave function $\psi_{\vep}(x)$ in the case of $\vf=0$ and  $\langle \hat{n}\rangle = 25 $,
compared with the coherent state wave function $\psi_{\alpha}(x)$ with $\alpha=5$.
Right: The probability density $|\psi_{\vep}(x)|^2$ for $\vf=0$ and  $\langle \hat{n}\rangle = 25 $, compared with
the Gaussian probability density (\ref{rhoGxx}) with $\langle x \rangle = 6.3$ and $\sigma_x=10$.
Series (\ref{psi-gen}) was calculated with $10000$ terms.
}}
\label{fig-psi-vf0}
\end{figure}

\begin{figure}[hbt]
\centering
\includegraphics[scale=0.11]{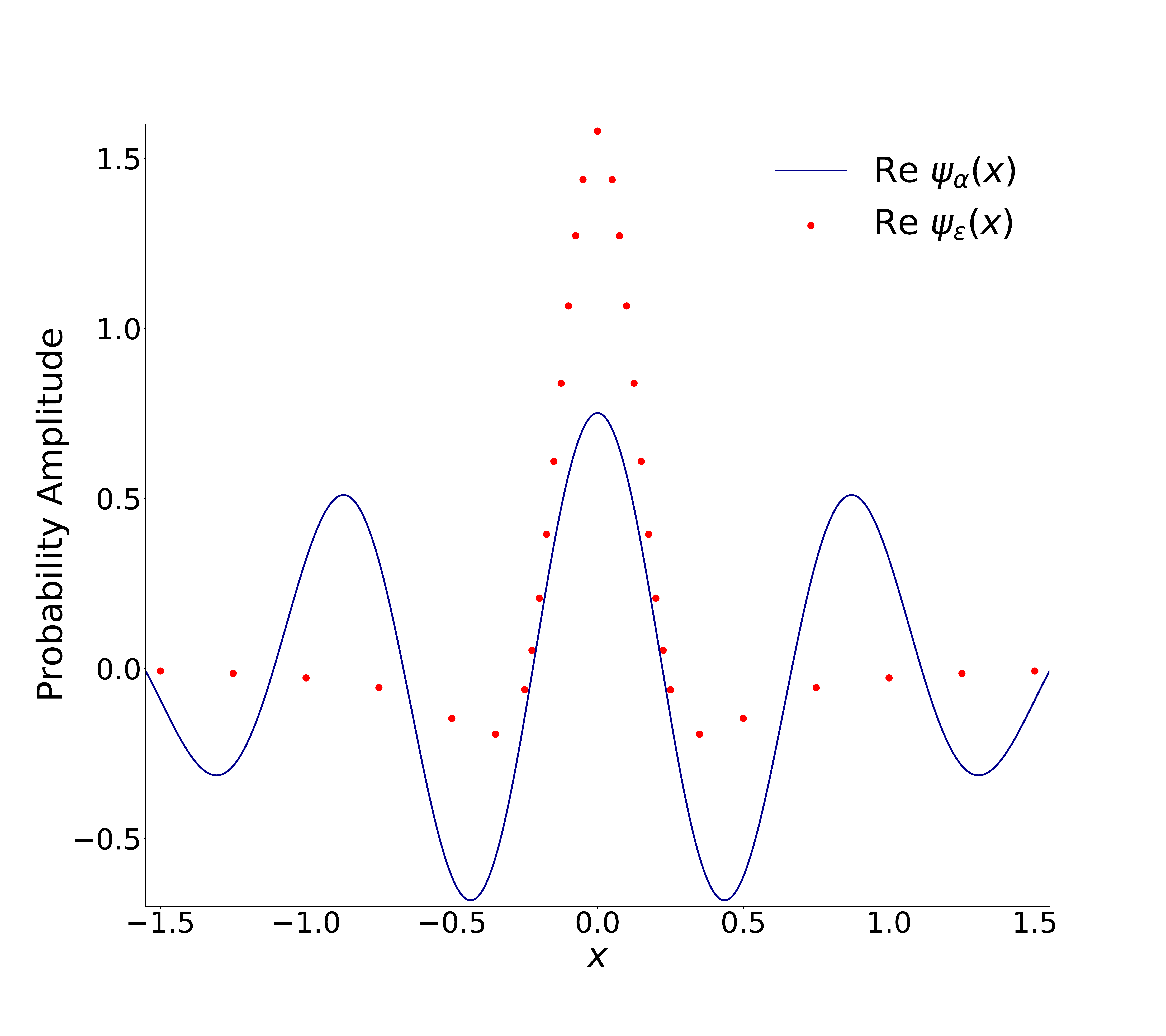}
\hfill
\includegraphics[scale=0.11]{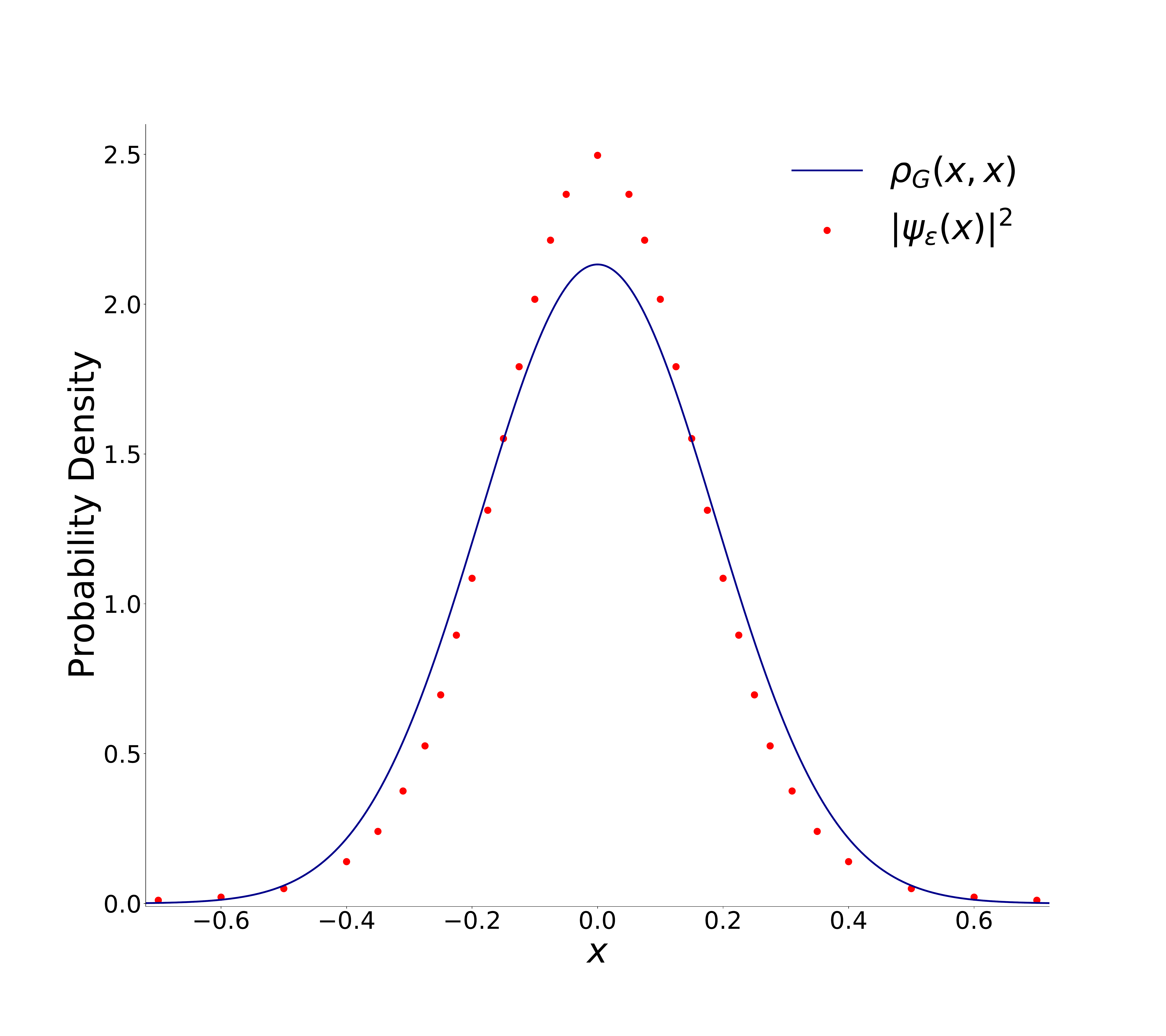}
\caption{\small{ Left: The real part of the wave function $\psi_{\vep}(x)$ in the case of $\vf= \pi/2$ and  $\langle \hat{n}\rangle = 25 $,
compared with the real part of the coherent state wave function $\psi_{\alpha}(x)$ with $\alpha=5i$, i.e.,
$\psi(x) = \pi^{-1/4}\exp(-x^2/2)\cos(5\sqrt{2}\,x)$.
Right: 
The probability density $|\psi_{\vep}(x)|^2$ in the case of $\vf= \pi/2$ and  $\langle \hat{n}\rangle = 25 $,
compared with the probability density $\rho_G(x,x)$
of the Gaussian state with the same values of the coordinate mean value $\langle x\rangle=0$ and variance $\sigma_x=0.035$.
Series (\ref{psi-gen}) was calculated with $10000$ terms.
}}
\label{fig-psi-pi2}
\end{figure}

Illustrations are given in Figures  \ref{fig-psi-vf0} and \ref{fig-psi-pi2}. 
In the left part of Figure  \ref{fig-psi-vf0} we compare the real wave function $\psi_{\vep}(x)$ 
with $\vf=0$ and  $\langle \hat{n}\rangle = 25 $ (i.e., $\vep=\sqrt{25/26}$)
with the real coherent state wave function $\psi_{\alpha}(x)$ (\ref{psicoh}) with $\alpha =5$ (i.e., having 
the same mean number of quanta $\langle \hat{n}\rangle = 25 $).
 The numeric summation in formula (\ref{psi-gen}) was performed for $0 \le n \le 1000$.
In the right-hand part of the same figure we compare the probability density $|\psi_{\vep}(x)|^2$ with
the probability density of the  Gaussian state, 
\be
\rho_G(x,x) = (2\pi\sigma_x)^{-1/2}\exp\left[- \left(x - \langle x \rangle \right)^2/(2\sigma_x)\right],
\label{rhoGxx}
\ee
having the same mean value of the coordinate $\langle x \rangle = 6.3$
and the same value of the variance  $\sigma_x=10$. 

Remember that the width of the coherent state wave function $\psi_{\alpha}(x)$ 
does not depend on the shift $\alpha$
(as soon as $\sigma_x \equiv 1/2$ in this case). In addition, the maximum of the coherent state wave function is achieved
at the point $x_m=\langle x \rangle =\sqrt{2}\,\alpha$, and the function is symmetric with respect to this point.
On the contrary, the function $\psi_{\vep}(x)$ shows a strong asymmetry with respect to the point of its maximum $x_m \approx 5 < \langle x \rangle \approx 6.3$, 
 and its width is significantly greater than that of the coherent state.

In Figure \ref{fig-psi-pi2} we plot similar functions for $\langle \hat{n}\rangle = 25 $, 
but now with $\vf= \pi/2$ and
$\alpha = 5i$. The reference Gaussian state in the right-hand side has the values
$\langle x\rangle =0$ and $\sigma_x^{\vf=\pi/2} = 0.035$.
In this figure, the difference between the real parts of wave functions of the usual coherent state and the phase coherent state
is quite impressive.

On the other hand, the right-hand parts of Figures \ref{fig-psi-vf0} and \ref{fig-psi-pi2} 
show that the {\em probability density\/}
profiles $|\psi_{\vep}(x)|^2$ seem visually {\em almost\/} as Gaussian distributions with the same coordinate mean values and
variances.
Therefore, it can be useful to have some quantitative measure of ``non-Gaussianity'' of the distribution $|\psi_{\vep}(x)|^2$.
Several existing measures were discussed, e.g., in Ref. \cite{CD21}. For a pure quantum state $|\psi\rangle$,
many of those measures are functions of the ``fidelity'' 
\be
F_G = \langle \psi|\hat\rho_G|\psi\rangle = \int\int dx dy \psi^*_{\vep}(x)\langle x|\hat\rho_G|y\rangle \psi_{\vep}(y),
\label{FG}
\ee 
where $\hat\rho_G$ is the statistical operator of the reference Gaussian state. 
It is natural to assume that the state $\hat\rho_G$ has the same mean values and variances of the quadrature components  
as the state $|\psi\rangle$ \cite{Gen07}. 
Unfortunately, the calculation of the double integral (\ref{FG}) meets 
strong computational difficulties, since the wave function $\psi(x)$ can be calculated only numerically and the density matrix
$\langle x|\hat\rho_G|y\rangle$ is not factorized for the mixed Gaussian state. 
Simple calculations can be done if one compares the
state (\ref{defCPS}) with its ``thermal analog'' (\ref{rhoth}), having the same value of parameter $\vep$. 
Then, the ``thermal fidelity'' $F_{th}= \langle \vep|\hat\rho_{th}|\vep\rangle$ can be easily found:
\be
F_{th}= \frac{1-|\vep|^2}{1+|\vep|^2} = \left(2\langle \hat{n}\rangle+1\right)^{-1}.
\label{Fth}
\ee
This quantity coincides with the ``purity'' $\mu = \mbox{Tr}\left(\hat\rho^2_{th}\right)$ of the thermal state (\ref{rhoth}).
However, the result (\ref{Fth}) seems unsatisfactory, because the probability densities $|\psi_{\vep}(x)|^2$ and
$\langle x|\hat\rho_{th}|x\rangle$ are very different for $|\vep|\to 1$, while the function
$|\psi_{\vep}(x)|^2$ with $\vf=0$ or $\vf=\pi/2$ looks  ``almost Gaussian'' 
in Figures \ref{fig-psi-vf0}-\ref{fig-psi-pi2}.

To find a more adequate and simple computable measure of ``Gaussianity'' we remember that all Gaussian probability distributions
in the coordinate space attain their maximal values at the point $x_*= 
\langle x \rangle$, where $|\psi(x_*)|^2 = (2\pi\sigma_x)^{-1/2}$.
Therefore, we use the quantity
\be
G = \sqrt{2\pi\sigma_x}\,|\psi(\langle x \rangle)|^2
\label{G}
\ee
as the measure of ``Gaussianity'' of pure quantum states. According to this definition, 
distributions can be called as
``superGaussian'' if $G>1$  and ``subGaussian'' if $G<1$. 

\begin{figure}[hbt]
\centering
\includegraphics[scale=0.11]{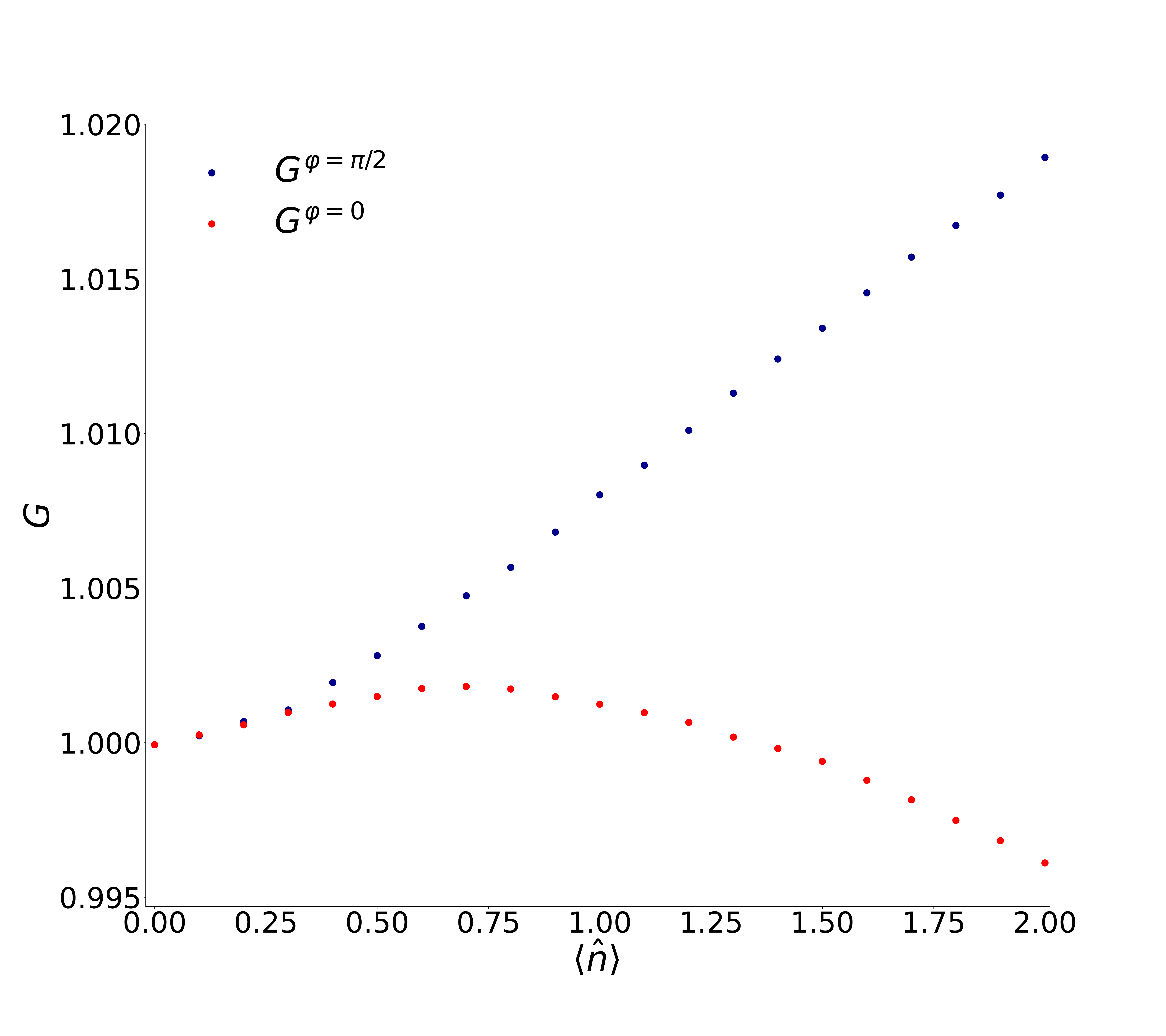}
\hfill
\includegraphics[scale=0.11]{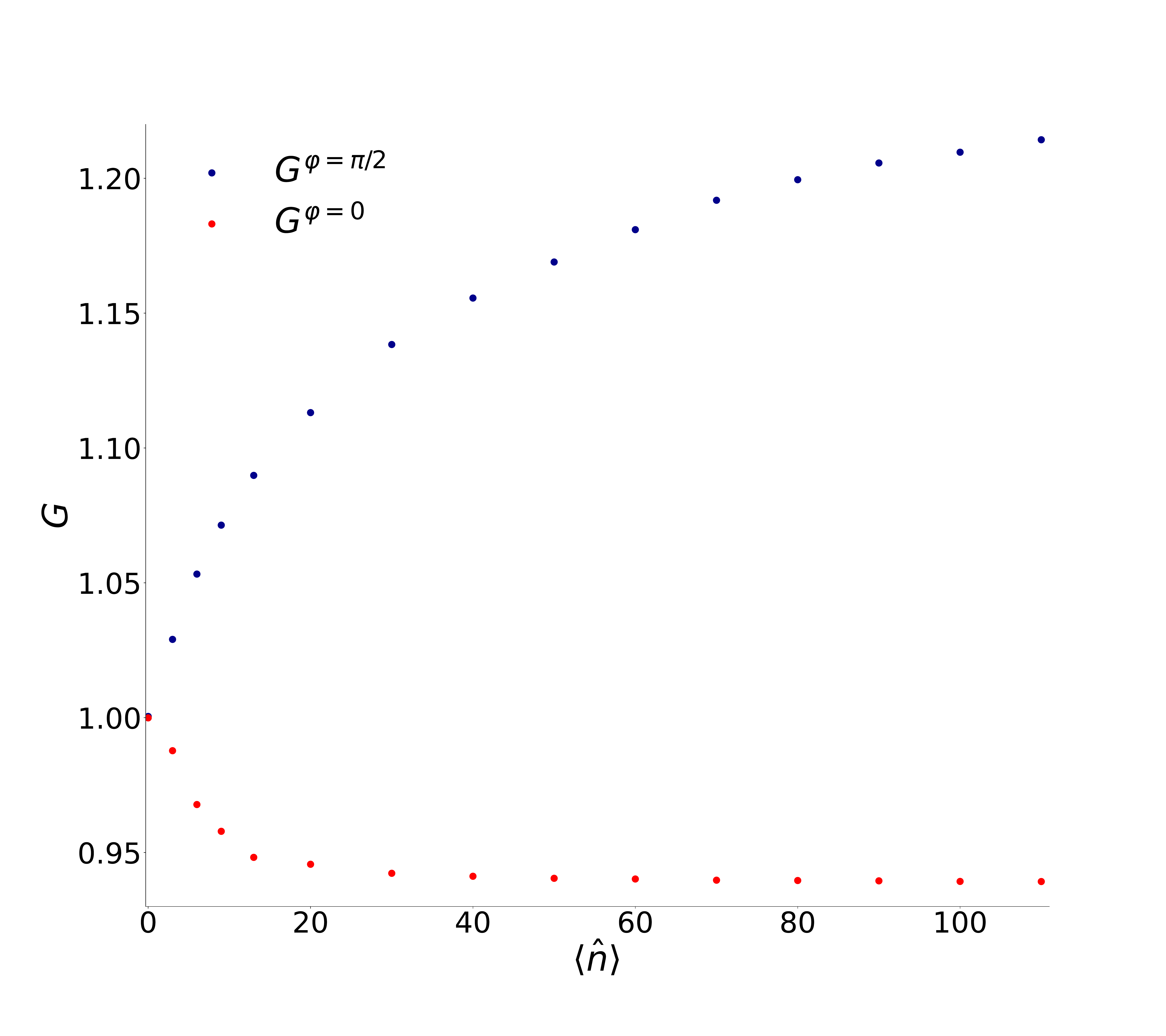}
\caption{\small{ The ``Gaussianity'' measure (\ref{G}) as function of $\langle \hat{n} \rangle$. Numeric calculations were performed
with $10000$ terms.
}}
\label{fig-G}
\end{figure}

The dependence $G(\vep)$ turns out non-trivial. In the limit of $\vep \to 0$,
taking into account only the terms of the order of $|\vep|$ and $|\vep|^2$,
we obtain $\langle x \rangle = \sqrt{2}\,\cos(\vf) S_1 \approx \sqrt{2}\,|\vep| \cos(\vf)$ and 
the following expressions:
\[
|\psi(\langle x \rangle)|^2 \approx \pi^{-1/2}\left[1 + |\vep|^2 \cos(2\vf)(1-\sqrt{2})\right], \qquad
\sqrt{2\pi\sigma_x} \approx \pi^{1/2}\left[1 + |\vep|^2 \cos(2\vf)(\sqrt{2}-1)\right].
\]
In this approximation, $G(\vep) \approx 1  + {\cal O}(|\vep|^4)$. Then, taking into account 
the terms of the order of $|\vep|^4$, we obtain the following (probably, unexpected) result
(see Appendix \ref{sec-ap-G} for details):
\be
G^{\vf=\pi/2}(\vep) =G^{\vf=0}(\vep) \approx 1 + |\vep|^4\left(3\sqrt{2} -3 -\sqrt{3/2}\right) + {\cal O}(|\vep|^6)
 \approx 1 +0.018|\vep|^4,
\label{Gsmall-pi2}
\ee
Hence, both these extreme states (with  $\vf=0$ and $\vf=\pi/2$) are ``weakly superGaussian'' if $\langle \hat{n}\rangle \ll 1$.
However, the behavior of two functions for larger values of $\langle \hat{n}\rangle$ is different, according to numeric calculations
whose results are shown in Figure \ref{fig-G}.

The most interesting behavior is observed for the dependence $G^{\vf =0}(\langle \hat{n}\rangle)$.
This function is slightly bigger than $1$ if $\langle \hat{n}\rangle < 1$. However, it shows the ``subGaussianity'' 
when $\langle \hat{n}\rangle \gg 1$. 
Unfortunately, we did not succeed in finding good analytic approximations for the functions $G^{\vf }(\langle \hat{n}\rangle)$.
Therefore, we do not know, whether $G^{\vf=0 }(\langle \hat{n}\rangle)$
goes asymptotically to zero or to some finite value when $\langle \hat{n}\rangle \to \infty$.
Also, we do not know, whether  $G^{\vf=\pi/2}$ grows slowly unlimitedly when $\langle n\rangle \to \infty$,
 or it approaches some asymptotic finite value.

\section{Wigner function}

In some cases, the ``non-Gaussianity'' can be seen distinctly in plots of the Wigner function
\be
W(q,p) = \int_{-\infty}^{\infty} dv e^{-ipv} \psi^*(q-v/2)\psi(q+v/2),
\label{defW}
\ee
because $W(q,p)$ must be negative in some regions of the phase space for any {\em pure\/} non-Gaussian quantum state \cite{Hudson}.
The Wigner function of the superposition state (\ref{defCPS}) can be represented as the series over the Weyl--Wigner symbols $W_{mn}$
of the diadic operators $|m\rangle \langle n|$:
\be
W = \left(1-|\vep|^2\right) \sum_{m,n=0}^{\infty} |\vep|^{m+n} e^{i(m-n)\vf} W_{mn},
\label{Wgen}
\ee
where \cite{Groen46,BartMoy49,CahGla69,DM137}
\be
W_{mn}(q,p) = 2^{1 +\lambda/2} (-1)^{\mu} \sqrt{\frac{\mu!}{\nu!}} b^{\lambda} e^{-b^2 -i\chi(m-n)} 
L_{\mu}^{\lambda}(2b^2), 
\label{Wmn}
\ee
\[
q+ip = b e^{i\chi}, \quad b \ge 0, \quad \mu = \mbox{min}(m,n), \quad \nu = \mbox{max}(m,n), 
\quad \lambda = |m-n|.
\]
Here, $L_n^{\alpha}(z)$ is the associated Laguerre polynomial, defined as in book \cite{BE}.

Formulas similar to (\ref{Wgen}) and (\ref{Wmn}) were obtained in paper \cite{Herzog93} for the ``phase state''
\be
|\phi\rangle = (2\pi)^{-1/2} \sum_{n=0}^{\infty} e^{in\phi}|n\rangle.
\label{phist}
\ee
However, the state (\ref{phist}) is un-normalizable, and this can lead to some difficulties \cite{Carr68}. 
Besides, the authors of \cite{Herzog93}
were interested in the dependence of their Wigner function on phase $\phi$ only, 
and not in the dependence on the quadrature
variables $q,p$. 
The 3D plots of the Wigner function of the coherent phase state (\ref{defCPS}) 
were shown in paper \cite{Gerry09} for 
$\vep=0.3$ and $\vep=0.9$. In the first case, the function looks as a slightly shifted Gaussian hill, 
without any visible
negativity. In the second example, the distribution was clearly non-Gaussian, with one high and three low hills and negative deeps
between them. Unfortunately, it is impossible to evaluate the value of negative depth in that 3D figures.
 Visually, the negativity demonstrated there is not strong.
 
The combination of Equations (\ref{Wgen}) and (\ref{Wmn}) yields the decomposition of the Wigner function 
in the sum
of the ordinary and double series, $W=W_1 + W_2$, where
\be
W_1 = 2\left(1-|\vep|^2\right) e^{-b^2}  \sum_{n=0}^{\infty} \left(- |\vep|^{2}\right)^n L_{n}(2b^2),
\label{W1}
\ee
\be
W_2 = 4\left(1-|\vep|^2\right)
 e^{-b^2}\sum_{\mu =0}^{\infty} \sum_{\lambda =1}^{\infty}  \left(- |\vep|^{2}\right)^{\mu}
 (\sqrt{2}|\vep| b)^{\lambda}
\cos[\lambda(\vf -\chi)] \sqrt{\frac{\mu!}{(\mu +\lambda)!}}\, 
 L_{\mu}^{\lambda}(2b^2).
\label{W2}
\ee
The ``diagonal'' sum $W_1$ is nothing but the Wigner function of the thermal state (\ref{rhoth}). 
Indeed, the known generating function of the Laguerre polynomials \cite{BE},
\[
\sum_{n=0}^{\infty} L_n(x) z^n = (1-z)^{-1} \exp\left(\frac{xz}{z-1}\right),
\]
results in the Gaussian distribution
\be
W_1 = \frac{2}{1+2\langle \hat{n} \rangle}\exp\left( -\,\frac{q^2 + p^2}{1+2\langle \hat{n} \rangle}\right).
\label{W1ans}
\ee
\begin{figure}[hbt]
\centering
\includegraphics[scale=0.11]{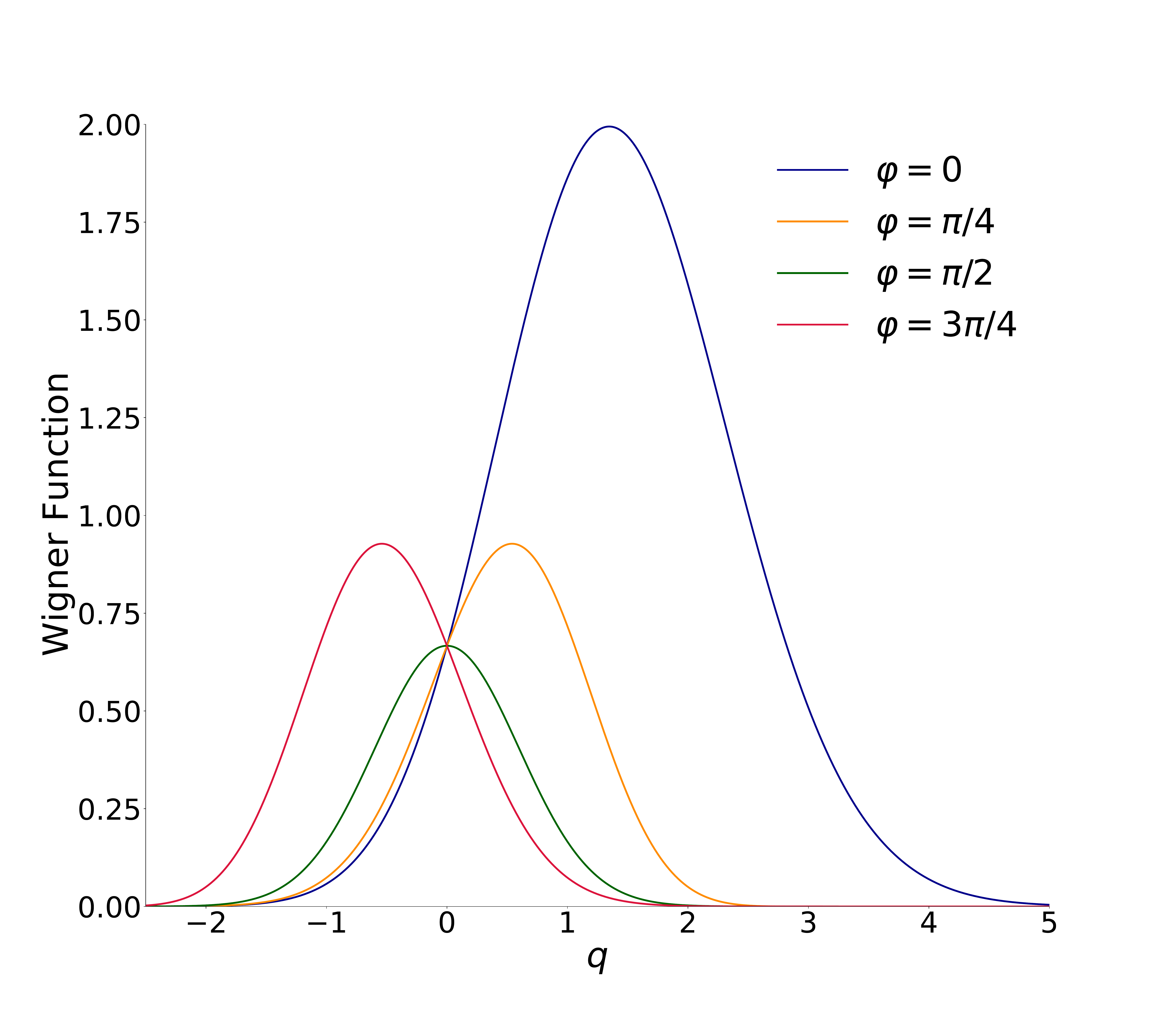}
\hfill
\includegraphics[scale=0.11]{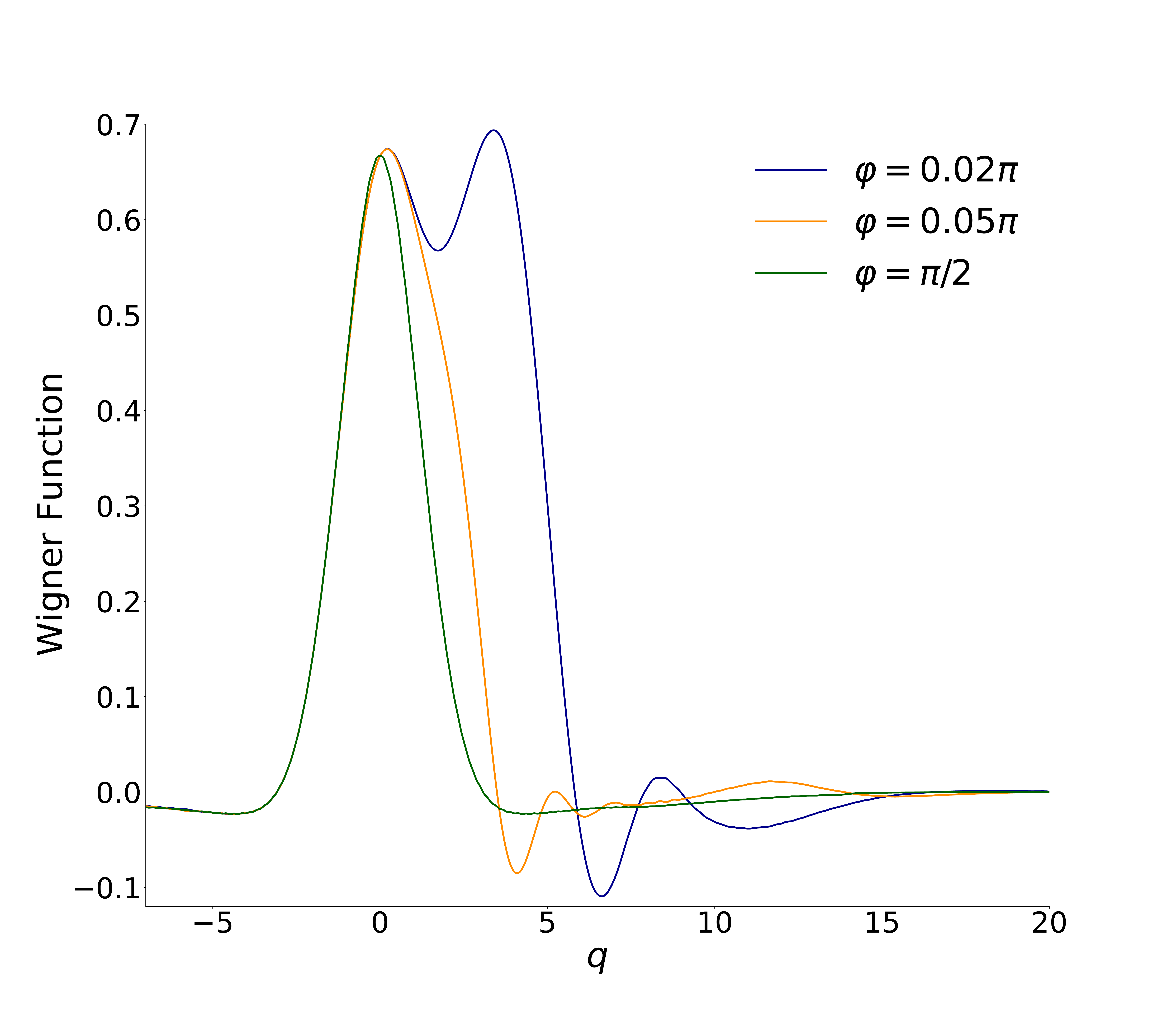}
\caption{\small{ 
2D sections of the Wigner function $W(q,0)$ for some values of the  phase $\vf$.
Left: for $\langle \hat{n} \rangle =1$. Right: for $\langle \hat{n} \rangle =30$.
}}
\label{fig-Wig}
\end{figure} 

Figure \ref{fig-Wig}  shows the 2D section of the Wigner function, $W(q,0; \vf)$, 
for different values of the phase $\vf$ and the mean quantum number  $\langle n \rangle$.
Plots of other 2D sections, such as $W(0,p)$ for example, look similar, 
as soon as they can be obtained from
$W(q,0; \vf)$ by means of some rotations in the phase plane $(q,p)$.
The summation in formula (\ref{W2})  was performed for $\mu \le 110$ and $\lambda \le 110$
in the case of $\langle n \rangle =30$.
Increasing the value of $\langle n \rangle$, we see how the initial Gaussian shape becomes 
more and more deformed, and negative values of the Wigner function become more visible.
However, these negative values remain relatively small even for large values
of $\langle n \rangle$.

\section{Conclusions}

Results of preceding sections show that some properties of the coherent phase states (CPS) 
are very different from the properties of the usual Klauder--Glauber--Sudarshan coherent states (CS),
while some other properties are more or less similar.
The main difference is in the squeezing properties and the uncertainty products. While the variances
of the dimensionless quadrature components do not depend on the parameter $\alpha$ in the case of 
standard CS, these variances are very sensitive to the parameter $\vep$ in the case of CPS.
Namely, a strong squeezing can be observed for certain values of the phase of complex number $\vep$
when $|\vep|$ is close to unity. This squeezing is only twice weaker than for the ideal
vacuum squeezed states with the same mean number of photons.

Another difference is in the value of the Robertson--Schr\"odinger uncertainty product (\ref{RSD}).
This product attains the minimal possible value $1/4$ in all usual coherent states. On the other hand,
it grows slowly (but unlimitedly) with increase of $|\vep|$ (without any dependence 
on the phase $\vf$ of the complex number $\vep$), as shown in Equation (\ref{Das}).

In addition, the shapes of the wave functions $\psi_{\alpha}(x)$ and $\psi_{\vep}(x)$
are quite different when the mean photon number $\langle n \rangle$ is large enough, as shown
in Figures (\ref{fig-psi-vf0}) and (\ref{fig-psi-pi2}).
However, the shapes of the probability density $|\psi_{\vep}(x)|^2$ or the Wigner function
are not very far from the Gaussians, at least for not extremely high values of $\langle n \rangle$,
according to Figures \ref{fig-G} and \ref{fig-Wig}.
It would be interesting to know, whether the non-Gaussianity remains weak in the limit
$\langle n \rangle \to\infty$. The problem is the extremely slow convergence of series
(\ref{psi-gen}) and (\ref{W2}) when $|\vep| \to 1$. We used up to 10000 terms in numeric calculations
of the ``Gaussianity'' parameter $G$
with $\langle n \rangle \le 200$, but we did not succeed in performing reliable
calculations for substantially higher values of $\langle n \rangle$. 
Calculations of the Wigner function are even more involved: we used $110\times110$ 
terms in the double sum
(\ref{W2}) to obtain a reasonable accuracy for $W(q,0)$ in the case of $\langle n \rangle=30$;
it was difficult to go to higher values of $\langle n \rangle$.

\vspace{6pt}

{M.C.F. acknowledges the support by the CNPq grant no. 144461/2021-8.
V.V.D. acknowledges the partial support of the Brazilian funding agency 
Conselho Nacional de Desenvolvimento Cient\'{\i}fico e Tecnol\'ogico (CNPq). }

\appendix 

\section{Calculations of the Gaussianity}
\label{sec-ap-G}

If $\phi=0$, then 
\beqn
\langle x\rangle &=& \sqrt{2} S_1 = \sqrt{2}\,|\vep|\left(1-|\vep|^2\right)
\left[1 + \sqrt{2}\,|\vep|^2 + \sqrt{3}\,|\vep|^4 +\cdots  \right]
\nonumber \\
&=&
\sqrt{2}\,|\vep|\left[1 + (\sqrt{2}\,-1)|\vep|^2 + (\sqrt{3}\, -\sqrt{2})|\vep|^4 +\cdots
\right],
\eeqn
\be
S_1^2 = |\vep|^2\left[1 + 2(\sqrt{2}\,-1)|\vep|^2 + (3 + 2\sqrt{3}\, -4\sqrt{2})|\vep|^4 +\cdots \right],
\ee
\be
S_1^4 = |\vep|^4\left[1 + 4(\sqrt{2}\,-1)|\vep|^2 + \cdots \right],
\ee
 so that
\be
|\psi_{\vep}(\langle x\rangle)|^2 =
 \pi^{-1/2} \left(1-|\vep|^2\right)\exp(-2S_1^2) \left(\sum_{n=0}^{\infty} \frac{ |\vep|^n H_n(\sqrt{2} S_1)}{\sqrt{2^n n!}}\right)^2
 = \pi^{-1/2} EH^2,
\ee
\be
E =  \left(1-|\vep|^2\right)\left(1 -2S_1^2 +2S_1^4  +\cdots\right)
= 1 -3|\vep|^2 +|\vep|^4(8 -4\sqrt{2}) 
+\cdots
,
\ee
\beqn
H &=&  1 + \frac{|\vep|}{\sqrt{2}}\left(2\sqrt{2} S_1\right) + \frac{|\vep|^2}{\sqrt{8}}\left[4(\sqrt{2} S_1)^2 -2\right]
 + \frac{|\vep|^3 (\sqrt{2} S_1)}{\sqrt{48}}\left[8(\sqrt{2} S_1)^2 -12\right]
 \nonumber \\
&+& \frac{|\vep|^4}{\sqrt{384}}\left[16(\sqrt{2} S_1)^4 -48(\sqrt{2} S_1)^2 +12\right] 
 +\cdots
  \nonumber \\
  &=& 1 + 2|\vep| S_1 - \frac{|\vep|^2}{\sqrt{2}} +\sqrt{8} (|\vep| S_1)^2 -\sqrt{6} |\vep|^3 S_1
  + \sqrt{\frac38}\,|\vep|^4  
 +\cdots 
     \nonumber \\
 &=& 1 +|\vep|^2 \left(2 -\frac12\sqrt{2}\right) +|\vep|^4 \left(4\sqrt{2}\, -2 - \sqrt{6}\,  +\sqrt{\frac38}\right)
 +\cdots 
\eeqn 
\be
H^2 = 1 +|\vep|^2 \left(4 -\sqrt{2}\right) +|\vep|^4 \left(6\sqrt{2}\, +\frac{1}{2} - 2\sqrt{6}\,  +\sqrt{\frac32}\right)
 +\cdots 
\ee
\be
EH^2 = 1 +|\vep|^2 \left(1 -\sqrt{2}\right) +|\vep|^4 \left(5\sqrt{2}\, -\frac{7}{2} - 3\sqrt{\frac32}\right)
 +\cdots 
\ee
\beqn 
\sigma_x &=& \frac12 + \frac{|\vep|^2}{1-|\vep|^2} -2S_1^2 +S_2 
  \nonumber \\
&=& 
\frac12 + |\vep|^2 + |\vep|^4  -2|\vep|^2\left[1 + 2(\sqrt{2}\,-1)|\vep|^2   \right] 
+ |\vep|^2(1-|\vep|^2)\left(\sqrt{2}\, + \sqrt{6}\,|\vep|^2 \right) +\cdots
  \nonumber \\
&=& \frac12 + |\vep|^2\left(\sqrt{2}\,-1\right) + |\vep|^4\left( 5 - 5\sqrt{2}\, +\sqrt{6}\right)
+\cdots
\eeqn
\be
\sqrt{2\sigma_x} = 1 + |\vep|^2\left(\sqrt{2}\,-1\right) + |\vep|^4\left( \frac72 -4\sqrt{2}\, +\sqrt{6}\right)
+\cdots
\ee
\be
G^{\phi=0}(\vep) = 
1 + |\vep|^4\left( 3\sqrt{2}\, -3 -\sqrt{\frac32}\right) +\cdots
\approx
1 + 0.018|\vep|^4  + \cdots
\ee

If $\vf = \pi/2$, 
\beqn
\pi^{1/2}|\psi(0)|^2 &=& \left(1- |\vep|^2\right) \left(1 + \frac{|\vep|^2}{\sqrt{2} } + |\vep|^4\sqrt{\frac38}\, 
  \right)^2 +\cdots
  \nonumber \\
&=&
  1 +(\sqrt{2}\, -1)|\vep|^2 +|\vep|^4\left( \frac12 + \sqrt{\frac32}\, -\sqrt{2}\right)
+\cdots
\eeqn
\be
\sigma_x = N - S_2 = \frac12 +(1-\sqrt{2})|\vep|^2 +|\vep|^4\left( 1 +\sqrt{2}\, -\sqrt{6}\right)
+\cdots
\ee
\be
\sqrt{2\sigma_x} = 1 +(1-\sqrt{2})|\vep|^2 +|\vep|^4\left( 2\sqrt{2}\, -\frac12 -\sqrt{6}\right)
+\cdots
\ee
\be
G^{\phi=\pi/2}(\vep) = 
1 + |\vep|^4\left( 3\sqrt{2}\, -3 -\sqrt{\frac32}\right)
+\cdots
\approx
1 + 0.018|\vep|^4  + \cdots
\ee

\end{document}